\documentclass[fleqn,usenatbib]{mnras}
\usepackage{newtxtext,newtxmath}
\usepackage[T1]{fontenc}
\usepackage{subfigure} 
\usepackage{graphicx}
\DeclareRobustCommand{\VAN}[3]{#2}
\let\VANthebibliography\thebibliography
\def\thebibliography{\DeclareRobustCommand{\VAN}[3]{##3}\VANthebibliography}

\usepackage{ulem}
\usepackage{CJKulem}
\usepackage{soul}

\title[Polarized radio emission of RRAT J1854+0306]{Polarized radio emission of RRAT J1854+0306}

\author[Qi Guo et al.]{
Qi Guo,$^{1,2}$
Minzhi Kong,$^{1,2}$\thanks{E-mail:kmz@hebtu.edu.cn}
P. F. Wang,$^{3,4}$\thanks{E-mail: pfwang@nao.cas.cn}    
Y. Yan,$^{3,4}$
D. J. Zhou$^{3,4}$
\\
$^{1}$College of Physics, Hebei Normal University, Shijiazhuang 050024, People’s Republic of China\\
$^{2}$Guo Shoujing Institute for Astronomy, Hebei Normal University, Shijiazhuang 050024, China\\
$^{3}$National Astronomical Observatories, Chinese Academy of Sciences, 20A Datun Road, Chaoyang District, Beijing 100012, China;\\
$^{4}$School of Astronomy, University of Chinese Academy of Sciences, Beijing 100049, China;
}

\date{Accepted: 04 April 2024.}

\pubyear{2024}

\begin{document}
\label{firstpage}
\pagerange{\pageref{firstpage}--\pageref{lastpage}}
\maketitle

\begin{abstract}
Polarized radio emission of RRAT J1854+0306 is investigated with single pulses using FAST. Its emission is characterized by nulls, narrow and weak pulses and occasional wide and intense bursts with a nulling fraction of 53.2\%. Its burst emission is typically of one rotation, and occasionally of two or three or even 5 rotations at the most, but without significant periodicity. The integrated pulse profile has a `S' shaped position angle curve that is superposed with orthogonal modes, from which geometry parameters are obtained. Individual pulses exhibit diverse profile morphology with single, double or multiple peaks. The intensity and width of these pulses are highly correlated, and bright pulses generally have wide profiles with multiple peaks. These nulling behaviours, profile morphology and polarization demonstrate that RRAT has the same physical origins as the normal pulsars.

\end{abstract}

\begin{keywords}
pulsars: individual: PSR J1854+0306 --- polarization
\end{keywords}



\section{Introduction}

Rotating Radio Transients (RRATs) represent a distinct group of neutron stars, which were first found in 2006 through reprocessing the Parkes Multi-beam Pulsar Survey data for transient radio bursts \citep{mll+06}. They were generally found through a single pulse search method. About 300 RRATs have been discovered up to now \citep{Deneva+etal+2016,Cui+etal+2017,Patel+etal+2018,Tyul+etal+2018a,Tyul+etal+2018b,Logvinenko+etal+2020,Good+etal+2021,mll+21,Abhishek+etal+2022,Bezuidenhout+etal+2022,Samodurov+etal+2022,Tyul+etal+2022,Dong+etal+2023,Samodurov+etal+2023,Zhou+etal+2023}. They are characterised by rare repeating bursts at constant dispersion measures. The duration of radio bursts lasts for 2$ \sim $30ms, and the interval between the bursts ranges from minutes to hours. Some RRATs exhibit underlying periodicity ranging from 41.5ms to 7.7s \citep{Zhang+etal+2007,Abhishek+etal+2022}. The rate of change of pulse period has been measured for some sources, from which the magnetic field strength was inferred, which ranges from $10^{11}$Gs to $10^{13}$Gs \citep{Abhishek+etal+2022}. These features are similar to those of the normal pulsars \citep{Abhishek+etal+2022}.

Even though more RRATs have been discovered from various search programs, like the Galactic Plane Pulsar Snapshot (GPPS) survey etc \citep{mll+21,Zhou+etal+2023}, the physics responsible for its sporadic emission remains unclear. Some explanations relating to pulsar nulling have been proposed \citep{Burke-Spolaor+etal+2010}. It was argued that RRATs have the same emission mechanisms as pulsars and manifest complete turning off of emission in the null states \citep{Zhang+etal+2007}, or that their emission is too weak to be detected in the null states for RRATs like B0656+14 \citep{Weltevrede+etal+2006}. Recently, \citet{Zhou+etal+2023} proposed that most RRATs are simply weak pulsars with sparse strong pulses or extreme nulling pulsars, since 43 RRATs that have been found in other telescopes appear as normal pulsars, extremely nulling pulsars, and weak pulsars with sparse strong pulses in Five-hundred-meter Aperture Spherical Telescope (FAST) observations. 

\begin{figure*}
        \centering
        \includegraphics[width=0.33\textwidth]{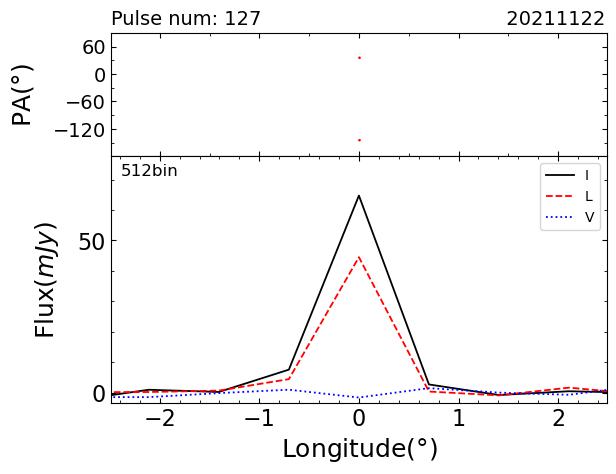}
        \includegraphics[width=0.33\textwidth]{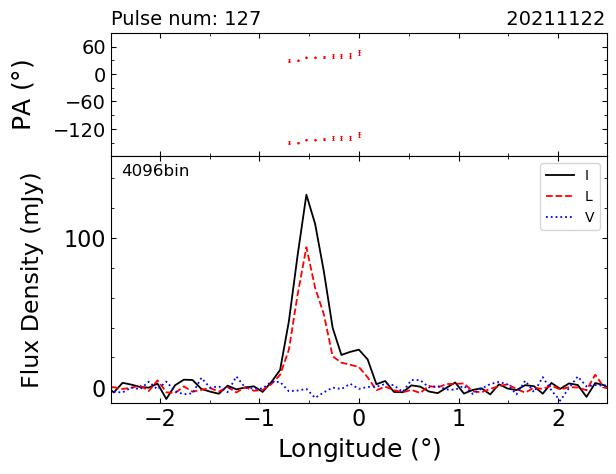}
        \includegraphics[width=0.33\textwidth]{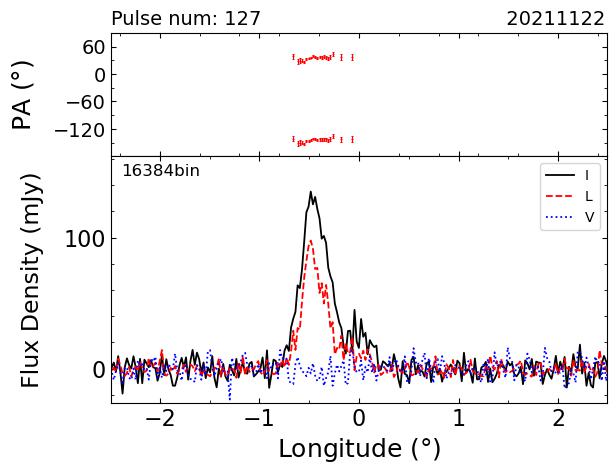}\\

   \caption{Polarized pulse profiles  with different phase bins. From left to right, the 127th pulse observed on 20211122 is exhibited with 512, 4096 and 16384 phase bins. The total intensity($I$), linear ($L$) and circular polarization($V$) of each pulse is represented by black solid, red dashed and blue dotted lines in the bottom panel of each plot. The PAs of linear polarization exceeding 5$\sigma_{L}$ are indicated by red dots with error-bars of $\pm$ $\sigma_{PA}$  in the top panels of the plots. $\sigma_{L}$ and $\sigma_{PA}$ represent the uncertainties of L and PA.}    
   \label{Fig:Ex.DiffB}
\end{figure*}

Polarization plays an important part in understanding the emission of RRATs. It can help to determine emission geometry, discriminate polarized X or O modes, and ascertain the interaction between both modes \citep{Wang+etal+2014}. However, polarization measurement remains a challenge, as limited by the sensitivity of RRAT observations. The first polarization measurement on RRATs was carried out for J1819-1458 in 2009 \citep{Karastergiou+etal+2009}. They found that its linear polarization position angle (PA) exhibits an S-shaped curve that is superposed with two orthogonal jumps. Interaction of the orthogonal modes leads a low degree of linear polarization. 
Currently, polarization observations have been made for 82 RRATs \citep{Caleb+etal+2019,Meyers+etal+2019, mll+21,Xie+etal+2022, Hsu+etal+2023, Zhong+etal+2023, Zhou+etal+2023}. 
However, only J1819-1458, J0628+0909 and J1913+1330 were studied on individual pulses with polarization. It was found that the average degree of linear polarization reaches 40\%, and it can even reach 100\% for a single burst. To gain a better understanding of the emission of RRATs, it is necessary to conduct more sensitive single pulse observations.

Among all the RRATs studied, RRAT J1854+0306 is among the strongest ones, which makes it possible to explore the emission details. Moreover, although it is strong, its profile morphology and polarization features remain uncovered. These are crucial for determining the RRAT's emission mechanism and comparing its behaviour to that of normal pulsars. RRAT J1854+0306 has a spin period of 4.56 s and was discovered by \citet{Deneva+etal+2009}. With sensitive FAST observations, it was estimated to have a burst rate of 102$\pm 10 hr^{-1}$ \citep{Lu+etal+2019}, and was classified as a weak pulsar with occasional strong pulses \citep{Zhou+etal+2023}.

In this paper, we study the polarized emission of RRAT J1854+0306 with single pulses. This paper is organized as follows. Observation and data reduction for the source are introduced in Section 2. Our results are presented in Section 3. Discussion and conclusions are given in Section 4.

\section{Observations and data reduction}
\label{sect:Obs}

We made a 15-minute tracking observation of RRAT J1854+0306 on November 22, 2021 (hereafter 20211122) with FAST \citep{nan+2006}. The observation was conducted using the central beam of the 19-beam receiver with a frequency range of 1000 to 1500 MHz, which was divided into 2048 channels. Before and after the observation, a periodic square wave signal generated from the noise diodes was injected for calibration. Data from both the source and the calibrator were recorded with a sampling time of 49.152 µs and full polarization. We also retrieved the other observation, 20191225, for the source from the open data in the FAST released archive. It was also made with full polarization, except with 4096 channels and lasting for an hour. 

The data was first de-dispersed with a publicly available dispersion measure (DM) of 192.4\,$\rm pc\,cm^{-3}$ \citep{Keane+etal+2011}, but there remained DM delays across frequency channels. To refine the DM, top ten pulses with the maximum signal-to-noise ratios were selected. The DM for each pulse was determined with a method that maximizes the pulse structure \citep{Hessels+etal+2019}. By weighted averaging of them, a more precise DM of 197.13±0.09 $\rm pc~ cm^{-3}$
was obtained. The data was then de-dispersed to form single pulse sequences again by using $\rm DSPSR$ software package \citep{van+Straten+etal+2011}. The radio frequency interference affected frequency channels are removed by using $\rm PSRCHIVE$ software \citep{Hotan+etal+2004}. Polarization is then calibrated by following the procedures proposed in \citet{Wang+etal+2023}. 
After which, the pulses are integrated over frequency channels and their PAs are corrected to an infinite frequency to eliminate the influence of Faraday rotation as caused by ionosphere and interstellar medium \citep[e.g.][]{Force+etal+2015}. To get a flat baseline for each pulse, we performed linear least square fittings to the off-pulse regions of Stokes parameters, I, Q, U and V, and then subtracted these fitted lines for the entire pulse window. The linear polarization intensity for phase bin $i$ is calculated as $\rm L_i= \sqrt{\rm Q_i^{2}+\rm U_i^{2}}$, with position angle $\rm PA_i=\textstyle\frac{1}{2}arctan(\textstyle\frac{\rm U_i}{\rm Q_i})$. The bias of $L_i$ as caused by the quadrature summation is corrected by following \citet{Wang+etal+2023}, which is $L_{c,i}=\sqrt{|L_i^2-(\sigma_Q^2+\sigma_U^2)|}$ if $L_i^2>(\sigma_Q^2+\sigma_U^2)$, otherwise, $L_{c,i}=-\sqrt{|L_i^2-(\sigma_Q^2+\sigma_U^2)|}$ . Here, $\sigma_Q$ and $\sigma_U$ are
the standard deviation of the off-pulse Q and U data.
The uncertainties of $\sigma_{\rm L}$ and $\sigma_{\rm PA}$ are estimated from $\sigma_{\rm Q}$ and $\sigma_{\rm U}$ with error propagation.

\begin{figure}
    \centering
    \includegraphics[width=0.89\linewidth]{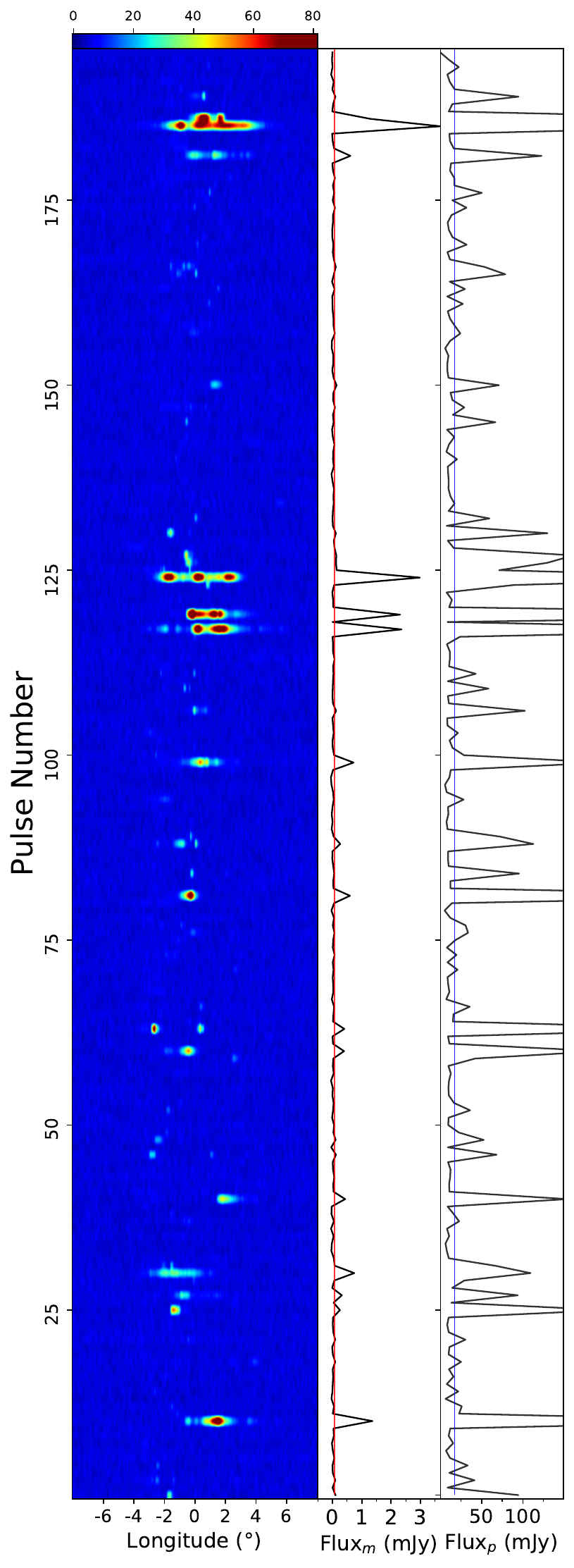}
\caption{Single pulse sequence observed on  20211122 with 196 pulse periods. {\it Left panel}: The pulse stacks with color indicating intensity. {\it Middle panel}: Flux density of individual pulses. The red line indicates the threshold of 3$\sigma_{\rm off}$, as estimated from the energy distribution in Figure~\ref{Fig:Energy}. {\it Right panel}: Peak intensity of individual pulses. The blue line indicates the threshold of 5$\sigma_{\rm bin}$, which is the root mean square of the off-pulse region for individual pulses.}
    \label{Fig:S.P.sequences}
\end{figure}

\begin{figure}
   \centering
   \includegraphics[width=1.0\columnwidth, angle=0]{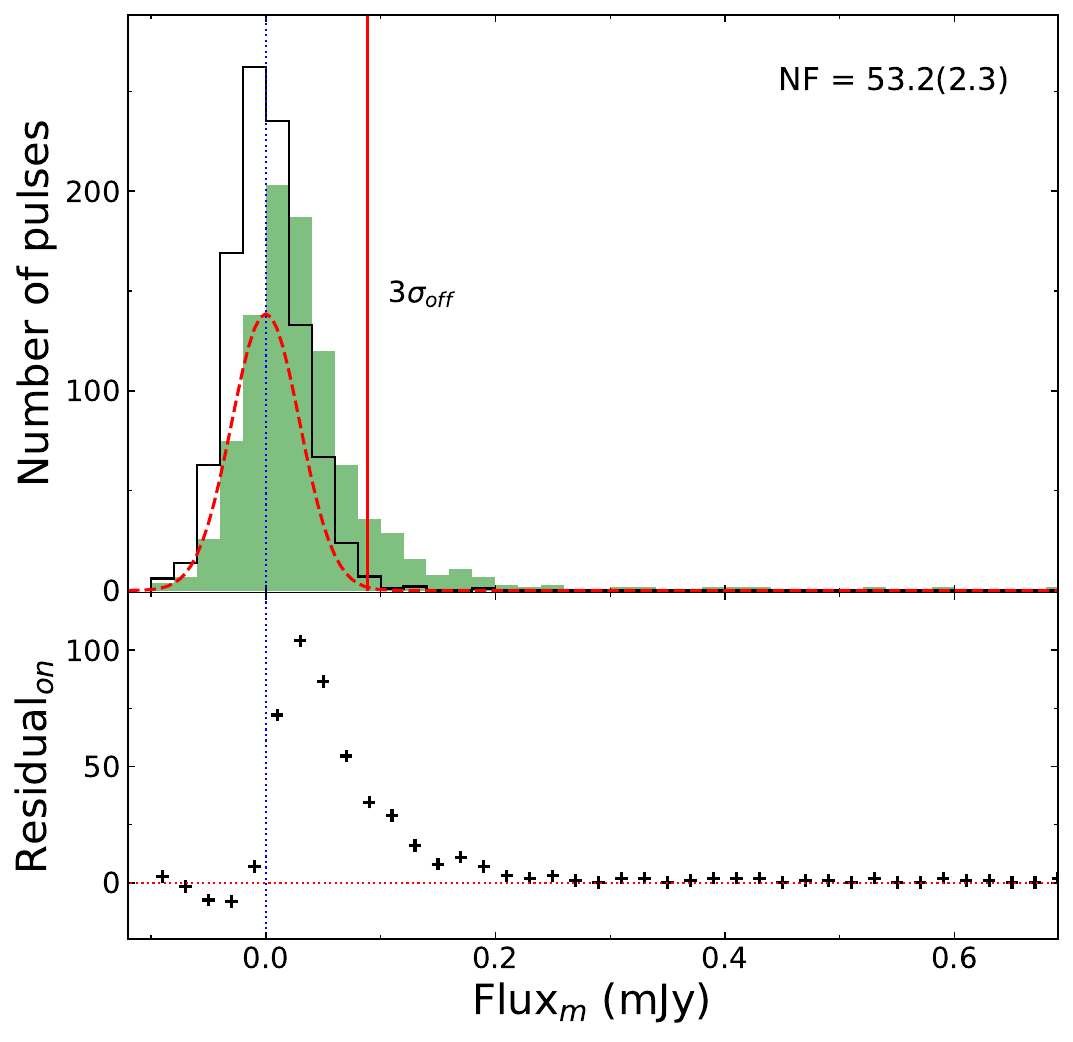}
   \caption{Histograms of the energy distributions for single pulses of J1854+0306 observed on 20191225 and 20211122 with 790 and 196 pulse periods. The on and off pulse distributions are indicated by green steps and no hatched black steps in the upper panel. The normal function representing the off-pulse distribution is scaled with nulling fraction (NF) and indicated by red dotted line. Residuals of the on-pulse distribution are indicated by black "+" dots in the bottom panel.}
   \label{Fig:Energy}
\end{figure}

\begin{figure}
    \centering

    \includegraphics[width=0.8\linewidth]{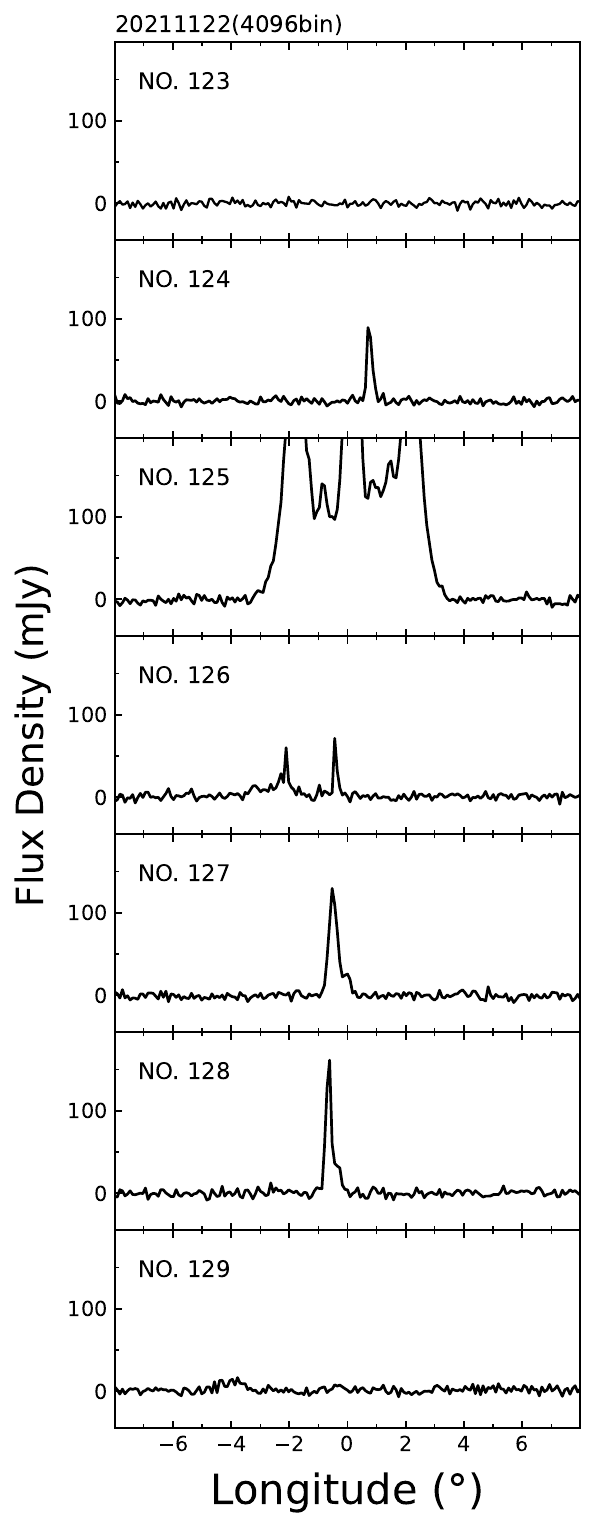}

   \caption{Pulse profiles for continuous individual pulses. The  sequence exhibits 5 continuous pulses observed on 20211122. The total intensity of each individual pulse is represented in each panel. Pulse number is indicated at the top left of panel and observation date is indicated at the top of plot.}
   \label{Fig:continuous}
\end{figure}

During our analysis, we noticed that time resolution of individual pulses affects their morphology classification and emission state identification. Comparison of the profiles are shown in Figure~\ref{Fig:Ex.DiffB} with 512, 4096 and 16384 phase bins. It is obvious that the No. 127 pulse observed on 20211122 appears a single peak in 512 bins, but which is resolved to exhibit multiple peaks in 4096 and 16384 bins. Furthermore, the pulse displayed at 4096bin appears to be smoother compared to the pulse displayed at 16384bin. All the analyses presented in this work are with 4096 bins.

\section{Results}

\subsection{Energy distribution}

Intensities of individual pulses vary greatly from one and the other. Their statistic is represented by the energy distribution. By which, the bursts, weak pulses and the nulls can be discriminated.
Single pulse sequence over the pulse window from observations are plotted in Figure~\ref{Fig:S.P.sequences}. Notably, changes of the pulse intensity and radiation width are noticeable in the pulse sequence. To get a more comprehensive understanding, we examine the energy distribution and emission states of these pulses.

The energy distribution and emission states of pulses are analyzed in the following. The on-pulse window is first chosen from the mean pulse profile, and the off-pulse window is selected from the off pulse regions with the same number of bins as the on pulse one. All the individual pulses have the same on and off pulse windows. The peak intensity represents the peak value of the bins in the on pulse window. The flux densities are calculated as the averages of the bin intensity within the on and off pulse windows.  They are binned to form histograms for their distributions, as presented in Figure~\ref{Fig:Energy} for all pulses from the two observations. The distribution of the off pulse flux densities can be modelled by a normal function with a mean of zero, an amplitude $A_0$, and a standard deviation of $\sigma_{\rm off}$. The normal function is again fitted to the on-pulse part with flux densities less than zero by fixing the mean and the so obtained $\sigma_{\rm off}$ to get $A_1$. The ratio of the amplitudes for the two normal distributions, $A_1/A_0$, represents the nulling fraction \citep{Wang+etal+2020}, which is estimated to be 53.2(2.3)$\%$.

It is apparent from Figure~\ref{Fig:Energy} that the peak location of the on-pulse distribution is far less than $3\sigma_{\rm off}$, which indicates the presence of many weak pulses, as manifested in Figure~\ref{Fig:S.P.sequences}. Although the pulses are generally weak, the occasional burst emission can reach a flux density of about 3 mJy. The threshold of 3$\sigma_{\rm off}$, i.e. about 0.09mJy, is selected to discriminate the bursts from the weak or null emission, as shown by the red lines in the middle panels of each sub-graph in Figure~\ref{Fig:S.P.sequences}. There are also a number of weak pulses whose flux densities are below the threshold, but having some intense bins far beyond the threshold. A new threshold, 5$\sigma_{\rm bin}$, is introduced to separate them, as shown by the blue lines in the right panels of each sub-graph in Figure~\ref{Fig:S.P.sequences}. Here, $\sigma_{\rm bin}$ represents the standard deviation of the total intensity for the off-pulses.  The so chosen pulses incorporate all the ones selected from the threshold of 3$\sigma_{\rm off}$. Finally, 319 bursts are identified from a total of 986 pulses with the threshold of 5$\sigma_{\rm bin}$. The ratio of nulling pulses is 67.6$\%$. It is larger than that estimated from the energy distribution, because only the pulses stronger than the threshold are identified here. But these values are within the wide range of nulling fractions of normal pulsars \citep{Wang+etal+2020}. 

\subsection{Burst duration and switching}

The physical processes responsible for the transitions between the null and burst states are still unclear. It has been argued that the short duration nulls are related to stochasticity in the emission process while the large-scale changes in the magnetosphere lead to the longer-duration nulls \citep{Cordes+etal+2013}. Whether the burst duration is correlated with the null duration and the underlying periodicity remain to be uncovered. 

The burst emission of J1854+0306 is typical of one rotation. Occasionally, the bursts last for two or three rotations, and even 5 rotations at the most. But it is very rare. Only two sequences exhibit 5 continuous pulses, and two exhibit 4 continuous pulses from two observations. As shown in Figure~\ref{Fig:continuous}, a sequence exhibits 5 continuous pulses observed on 20211122. 

Within the scope of energy store release mechanism \citep{Keane+etal+2011,Shapiro+etal+2018}, a pulsar stores its energy for the burst emission during the nulling state. Hence, the length of a null state might correlate with the released energy before or after the null. The periods that a pulsar stays in null is called waiting time. We correlate its length with the summed flux density of the bursts before or after the nulls. No correlation is found. It means that the burst emission of J1854+0306 is not generated by the energy store-release mechanism.

To check the periodicity of bursts, we first construct 18 sets of state  sequences with thresholds ranging from 3$\sigma_{\rm off}$ to 20$\sigma_{\rm off}$. Any pulse with intensity below or above the thresholds is denoted as of state 0 and 1, which represent null and burst state of the emission by following \citet{Wang+etal+2020}. A one-dimensional discrete Fourier transform (DFT) is then performed on these 0-1 sequences. Finally, no significant periodicity is found even with these diverse thresholds.

\subsection{Integrated pulse profile and geometry}
\begin{figure*}
    \centering  
    \includegraphics[width=0.33\linewidth]{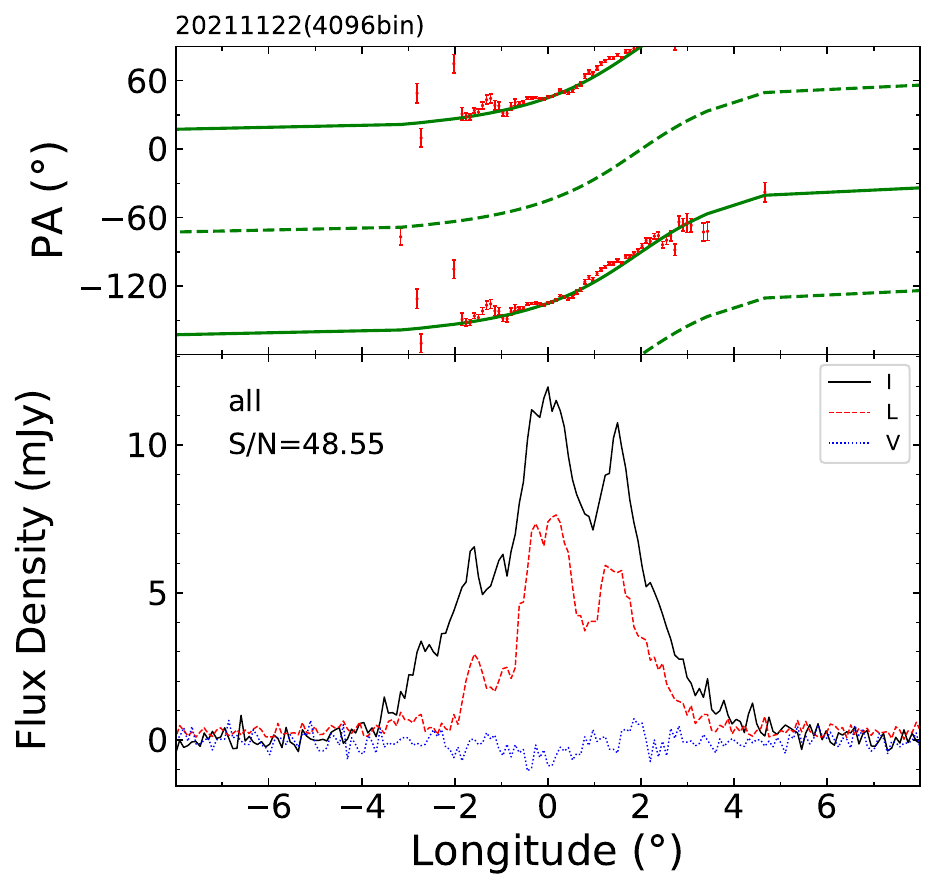}
    \includegraphics[width=0.33\linewidth]{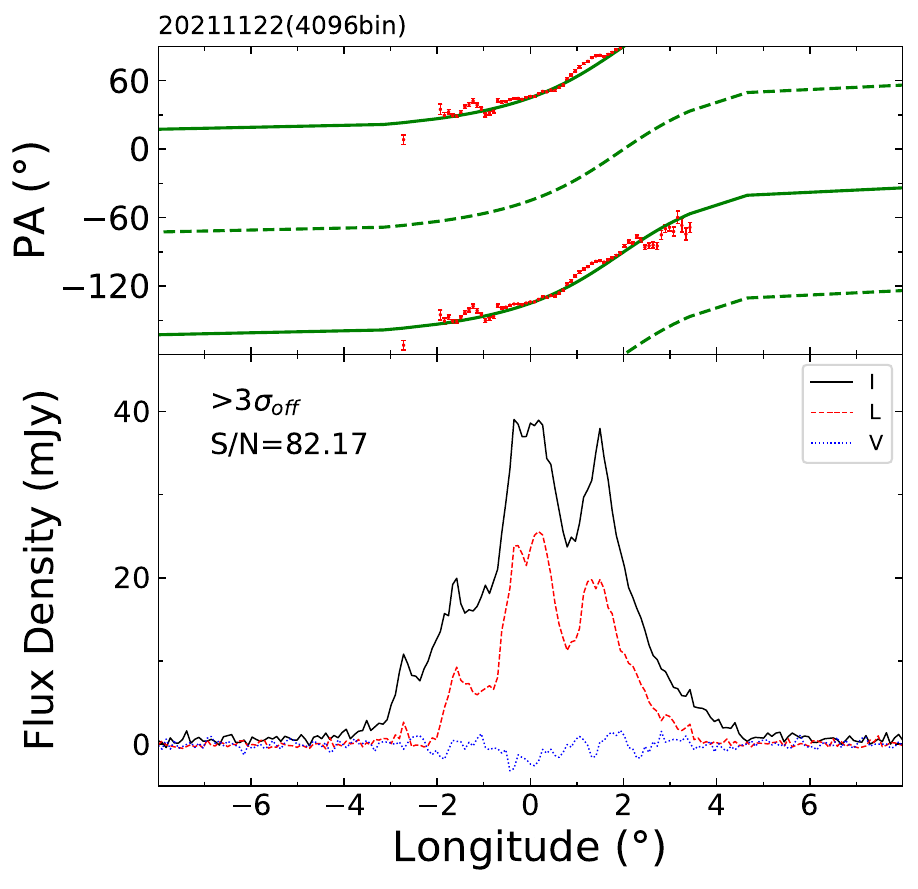}
    \includegraphics[width=0.33\linewidth,height=0.31\textwidth]{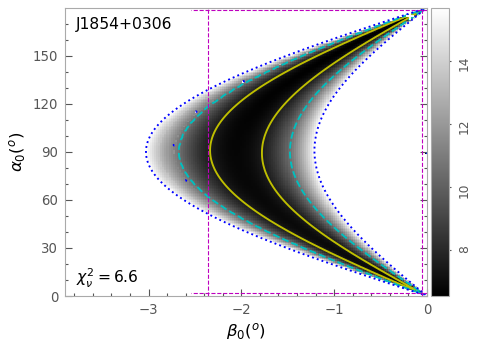}

   \caption{Integrated pulse profiles observed on 20211122. For this observation, {\it left plot:} integrated pulse profile of all pulses, {\it middle plot:} integrated profile of the pulses with flux density exceeding 3$\sigma_{\rm off}$. The total intensity, linear and circular polarization are plotted in the bottom panel of each plot. The linear polarization intensities exceed 5$\sigma_{L}$ for the plotted PAs. The PAs are indicated by red dots with error-bars in the top panels of the plots. Rotating vector model solution of the source is represented by the green solid and dashed lines separating by $90^\circ$. {\it right plot:}  $\chi^2$ surface of $\alpha_0$ and $\beta_0$ for the observations on 20211122. The minimum of $\chi^2$ is indicated by “+” with its value marked by $\chi^2_\nu$ in the panel. The solid, dashed and dotted contour lines are for $\Delta\chi^2$= 1.0, 4.0 and 9.0, respectively. The horizontal and vertical dashed pink lines indicate the projection of the solid contour line bounded region on the $\alpha_0$ and $\beta_0$ axes  \citep{Wang+etal+2023}.  }
\label{Fig:Integ}
\end{figure*}


The integrated pulse profile, together with its polarization, is the basis for understanding the emission processes in the pulsar magnetosphere.
Integrated pulse profile represents a unique feature of each pulsar. When a sufficient number of individual pulses are accumulated, the average profile of a pulsar becomes stable. Integrated pulse profiles are obtained for all the pulses and pulses with flux density $>3\sigma_{\rm off}$ from the observation on 20211122, as shown in Figure~\ref{Fig:Integ}. It is obvious that the average pulse profile extends about $9^\circ$ in rotation phase.

Its position angles generally follow an `S' shaped variation,  which can be described by the rotating vector model (RVM) \citep{Radhakrishnan+etal+1969}. RVM is widely employed to estimate the inclination angle of the magnetic axis with respect to the rotation axis and the impact angle of the sight line with respect to the magnetic axis, i.e., $\alpha_0$ and $\beta_0$. During the calculation, the full parameter spaces of $\alpha_0$ from $0^\circ$ to $180^\circ$ and $\beta_0$ from $-4^\circ$ to $0^\circ$ are explored to find the minimum of $\chi^2$ for the fittings by following \citet{Wang+etal+2023}. The inclination angle $\alpha_0$ is estimated to be of $175^{+4}_{-173}$ degree, the impact angle $\beta_0$ of $-0.18^{0.14}_{-2.18}$ degree, as shown by the modeled PA curves in Figure~\ref{Fig:Integ}. Here, the so obtained $\alpha_0$ and $\beta_0$ are highly covariant, as shown in the right plot of Figure~\ref{Fig:Integ}, and this solution is consistent with the one obtained by \citet{Wang+etal+2023} within the error limits. It indicates that the sight line cuts across its beam close to the magnetic axis. Its linear polarization degree is about 50$\%$, and circular polarization is nearly zero.

\subsection{Distribution of individual pulses}

\begin{figure}
   \centering \includegraphics[width=0.99\columnwidth]{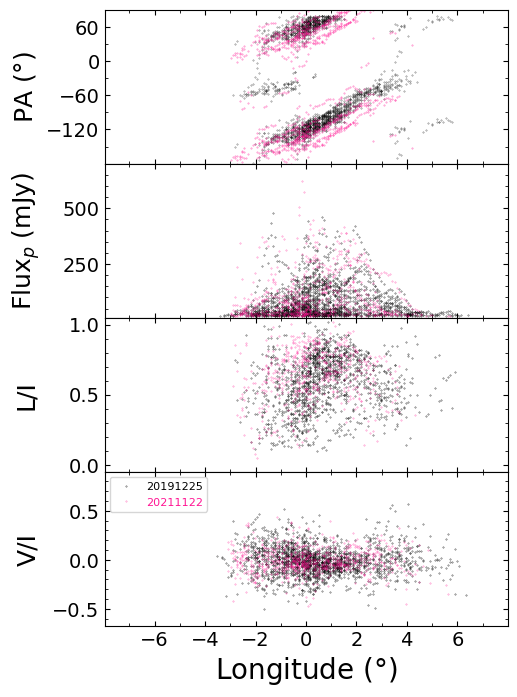}
    \caption{Distribution of individual pulses. From top to bottom, it shows the position angles, pulse intensities, degrees of linear and circular polarization($L/I$,$V/I$) of the single pulse bins with intensity exceeding the threshold of $5\sigma_{\rm bin}$. The linear polarization intensities also exceed 5$\sigma_{L}$ for the plotted PAs. The black and pink dots are observed on 20191225 and 20211122, respectively.}
\label{Fig:IndivDis}
\end{figure}

\begin{figure}
    \centering
    \includegraphics[width=0.48\textwidth]{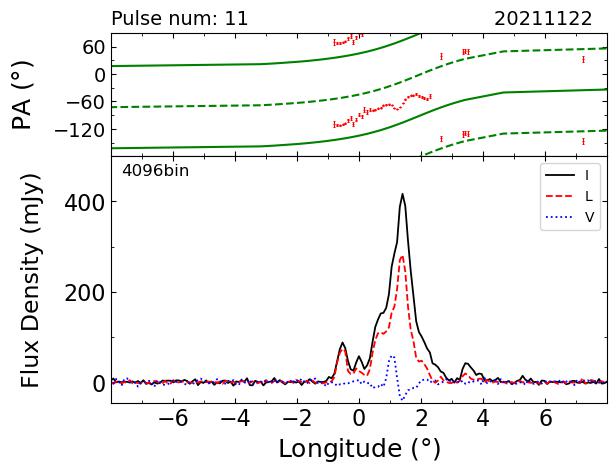}\\

    \includegraphics[width=0.48\textwidth]{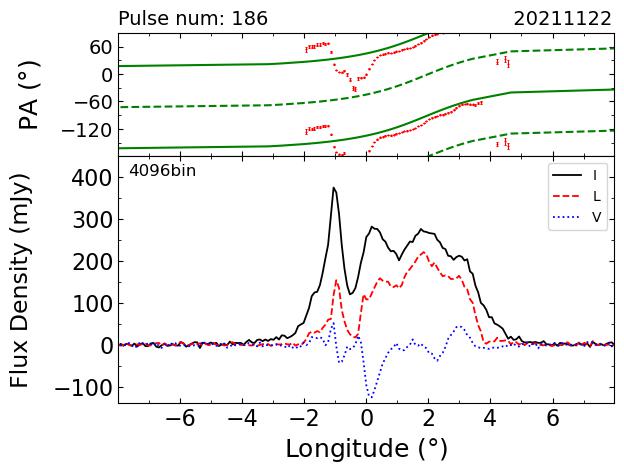}
   \caption{Polarized pulse profiles of 2 individual pulses whose position angles depart from the average PA curve. PAs of each individual pulse and average PA curve are shown in red with error bars and green lines in the top panels. The total intensity, linear and circular polarization are plotted in the bottom panels. The keys are the same as those in Figure~\ref{Fig:Ex.DiffB}.}
   \label{Fig:Ex.PA}
\end{figure}

Individual pulse studies are important to understand the pulsar emission mechanism.
The origin of orthogonal jumps that are difficult to explain in the average profile can be revealed through polarization of individual pulses. Unlike average pulse profiles, polarization of individual pulses varies a lot from one to another. Distribution of the polarization of individual pulses is shown in Figure~\ref{Fig:IndivDis} for two observations.  

It is apparent from the PA distribution that the pulses exhibit two distinct modes of polarization that separate by $90^\circ$.
Emission at the leading and trailing parts of pulse profiles is of one mode, and the emission at the central part is of the other. Orthogonal mode competition happens around the phases from -3.0$^\circ$ to 0$^\circ$, and from 3.5$^\circ$ to 4.0$^\circ$.

The change of their dominance occurs at the leading part of pulse profiles, which results in orthogonal mode jump as seen in Figure~\ref{Fig:Integ}(a). However, the dominance of the polarization modes remain unchanged in the observation on 20211122. Hence, no orthogonal mode jump is seen in Figure~\ref{Fig:Integ}(b). Due to the interaction of both modes, linear polarization is depolarized, and is of low fraction. These behaviors meet with the prediction of pulsar polarization emission theories that account for emission processes and propagation effects \citep{Wang+etal+2014}. According to this study, emission at the central phase of J1854+0306 is of X-mode, while the outer ones are of O-mode.

The PA distribution is much more dispersed for the observation on 2021112, which is the most significant for No. 11 and 186 pulses, as shown in Figure~\ref{Fig:Ex.PA}. The difference reaches about 30$^\circ$ for the No. 11 pulse. These differences can be caused by emission from different sets of relativistic particles. They might locate at different regions of pulsar magnetosphere, where the structure of magnetic field, their density and energy distributions are quite different \citep[e.g.][]{Philippov+etal+2020,Cruz+etal+2021}. The emission of these particles will experience different aberration \citep[e.g.][]{BCW91,Wang+etal+2012}, retardation \citep[e.g.][]{gan+05}, refraction \citep[e.g.][]{BA86}, and propagation effects \citep[e.g.][]{BP12,Wang+etal+2014}. These diverse influences result in the departure of polarization from those of the normal pulses.  

\subsection{Profile morphology of individual pulses}
\begin{figure*}
\subfigure[one peak]{
    \centering
    \includegraphics[width=0.32\textwidth]{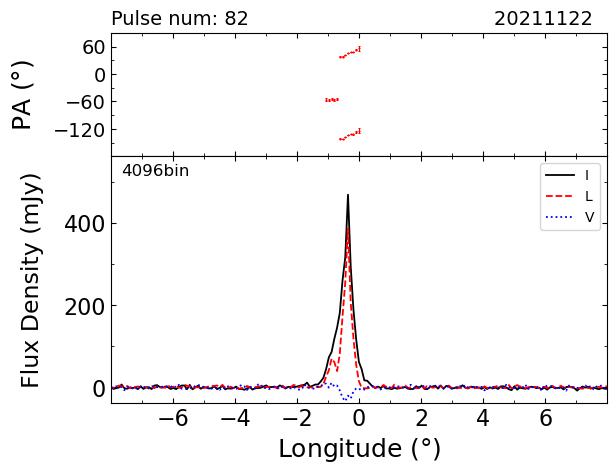}
}
\subfigure[Two peaks]{
    \centering
    \includegraphics[width=0.32\textwidth]{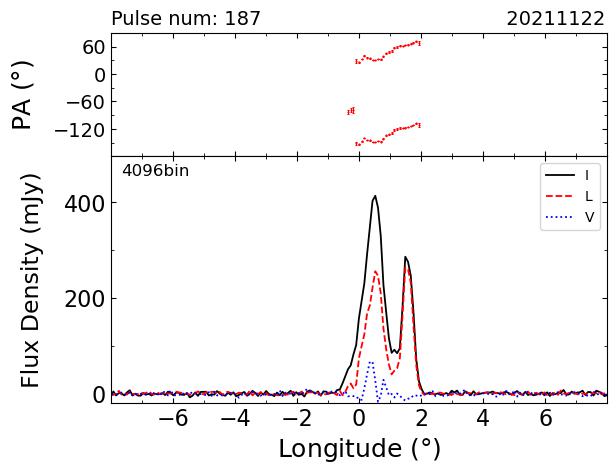}
}
\subfigure[Multiple peaks]{

    \includegraphics[width=0.32\textwidth]{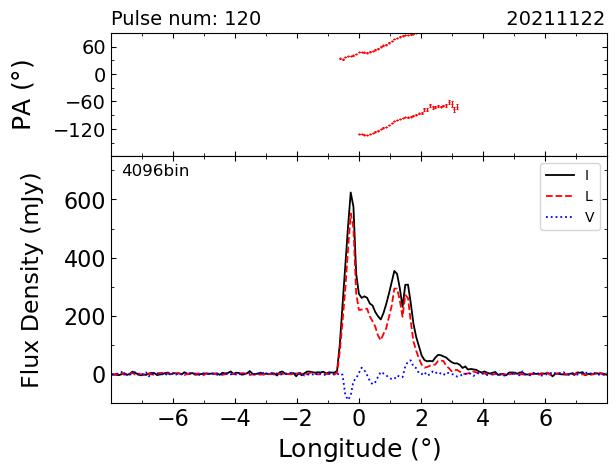}

}
  
   \caption{Examples of three types of individual pulse profiles. (a) Pulse profile with one peak. (b) Pulse profiles with two peaks. (c) Pulse profile with multi peaks. The keys are the same as those in Figure~\ref{Fig:Ex.PA}.}
   \label{Fig:Ex.Comp}
\end{figure*} 

\begin{table*}
\begin{center} 
 \caption{Statistics of three types of individual pluses from two observations.} 
 \label{tab:1}

\begin{tabular}{c|ccccc}

\hline
& No. &Peak Long. &Flux$_{\rm p}$ &Flux$_{\rm m}$& W\\
&&($^\circ$)& (mJy)& (mJy)& ($^\circ$)\\
(1)&(2)& (3)& (4)&(5)& (6)\\
\hline
\multicolumn{6}{c}{20191225}\\
\hline
One peak &148& -0.04[-2.90,5.71] &48.9[14.4,360.5]&0.07[-0.11,0.97]&0.22[0.09,1.93]\\ 
Two peaks & 48 & -0.09[-2.29,1.67]&55.6[16.68,197.90] &0.08[-0.10,0.40] & 0.74[0.28,3.08]\\
Multiple peaks & 51 & 0.73[-1.93,3.60]     &117.1[21.7,480.6] &0.43[-0.07,2.83] &2.57[0.28,7.29]\\
\hline
\multicolumn{6}{c}{20211122}\\
\hline
One peak & 45 & -0.49[-3.78,5.63] &54.8[15.3,469.2]  &0.07[-0.03,0.60] & 0.24[0.09,1.76] \\

Two peaks    & 13  & -0.70[-2.90,0.52]& 131.7[22.9,413.4] &0.21[0.01,1.29] & 1.53[0.46,4.21] \\

Multiple peaks & 14 & -0.34[-2.55,1.41] &221.5[29.2,624.0]&1.13[0.09,3.68] &3.6[0.54,7.38] \\    
\hline

\multicolumn{6}{l}{Note: Columns (2) to (6) are for the number of pulses, averages of their peak longitude, peak intensity, flux density and width. } \\
\multicolumn{6}{l}{The range of each parameter is indicated in the bracket after the value.} \\

\end{tabular}
\end{center} 
\end{table*}

\begin{figure*}
    \centering
    \subfigure[]{
    \includegraphics[width=0.326\linewidth]{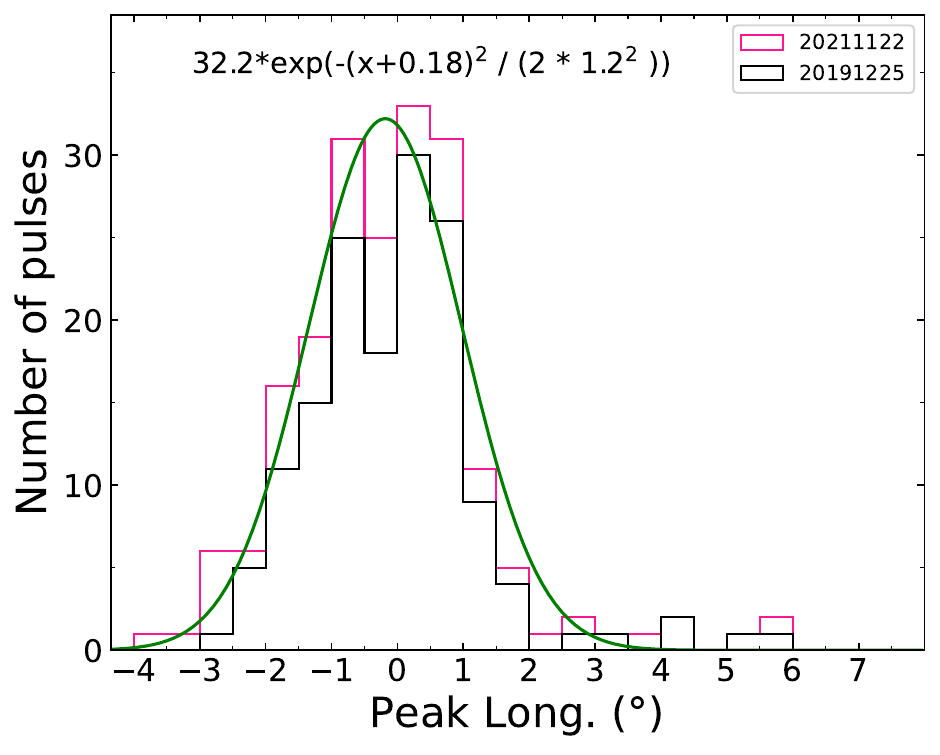}}   
    \subfigure[]{
    \includegraphics[width=0.326\linewidth]{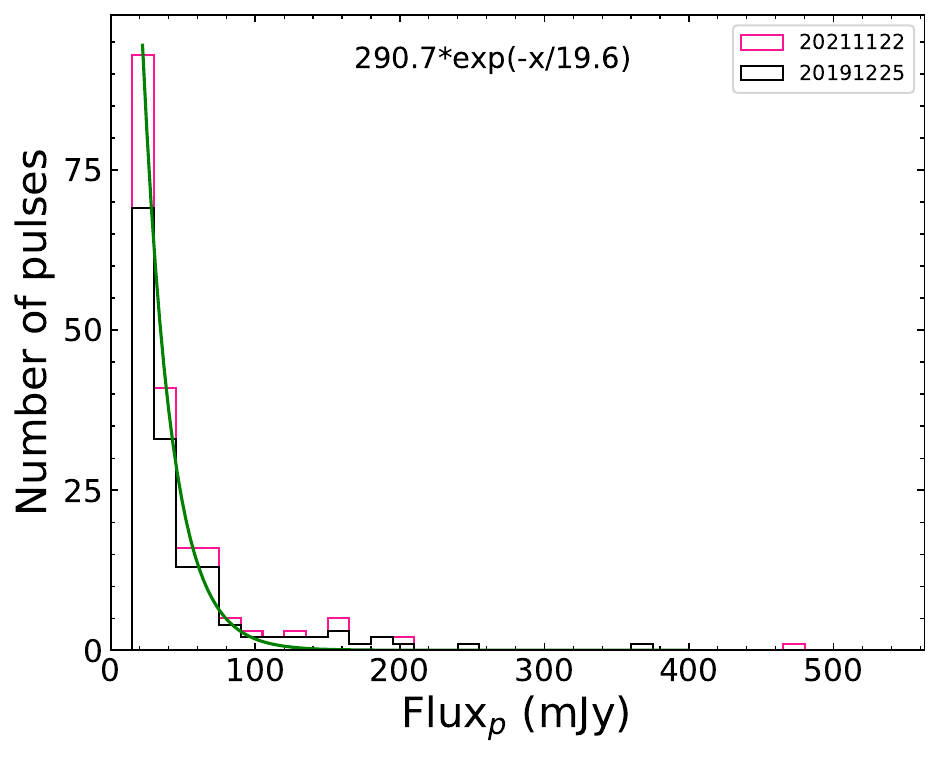}}
    \subfigure[]{
    \includegraphics[width=0.326\linewidth]{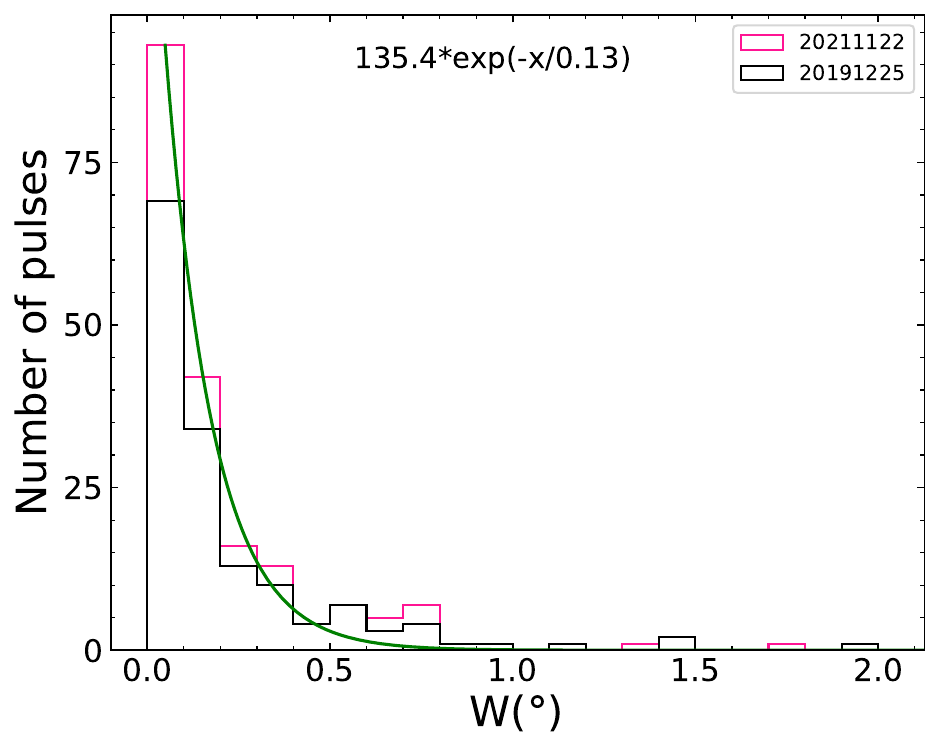}}

   \caption{Histograms of peak longitude, peak intensity and profile width for 193 pulses with one peak. 148 pulses are from the observation on 20191225 as indicated by black steps, and the pink steps are for 45 pulses from the observation on 20211122.}
   \label{Fig:OneComp}
\end{figure*}

A popular theory for the origins of pulsar radio emission is that it is generated by coherent radiation from discrete bundles of relativistic particles \citep{GHW21}. The profile shapes of individual pulses reveal the instantaneous distribution of relativistic particles. 319 individual pulses have peak intensities above the threshold of $5\sigma_{\rm bin}$ from the two observations. Their profiles exhibit diverse morphology, including single component, double peaks, multiple peaks.
To determine the component number, a pulse profile is first smoothed with one to several Gaussian functions. The component numbers are then confirmed by counting the peaks of the smoothed profiles. As introduced in section 2, the number of phase bins affects the determination peak number. For the current 4096 phase bins, any peak narrower than the bin size of about $0.09^\circ$ can not be resolved. 
Each type has different pulse numbers, peak longitudes, peak intensities, mean flux density and profile widths, as listed in columns (2) to (6) in Table~\ref{tab:1}. 
\subsubsection{Profiles with one component}
\begin{figure*}
    \centering
    \subfigure[peak intensity>5$\sigma_{\rm bin}$]{
    \includegraphics[width=0.49\textwidth]{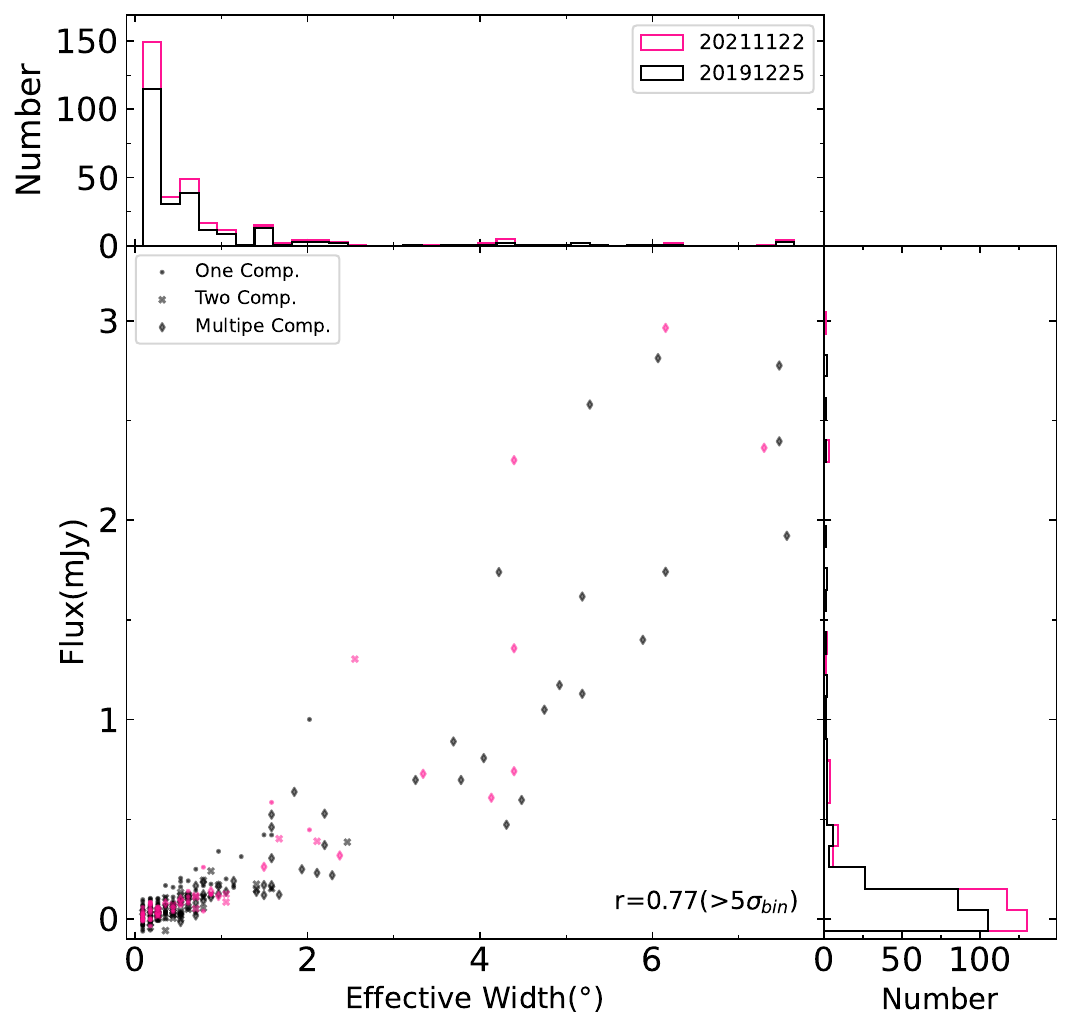}}   
    \subfigure[peak intensity>20$\sigma_{\rm bin}$]{

    \includegraphics[width=0.49\textwidth]{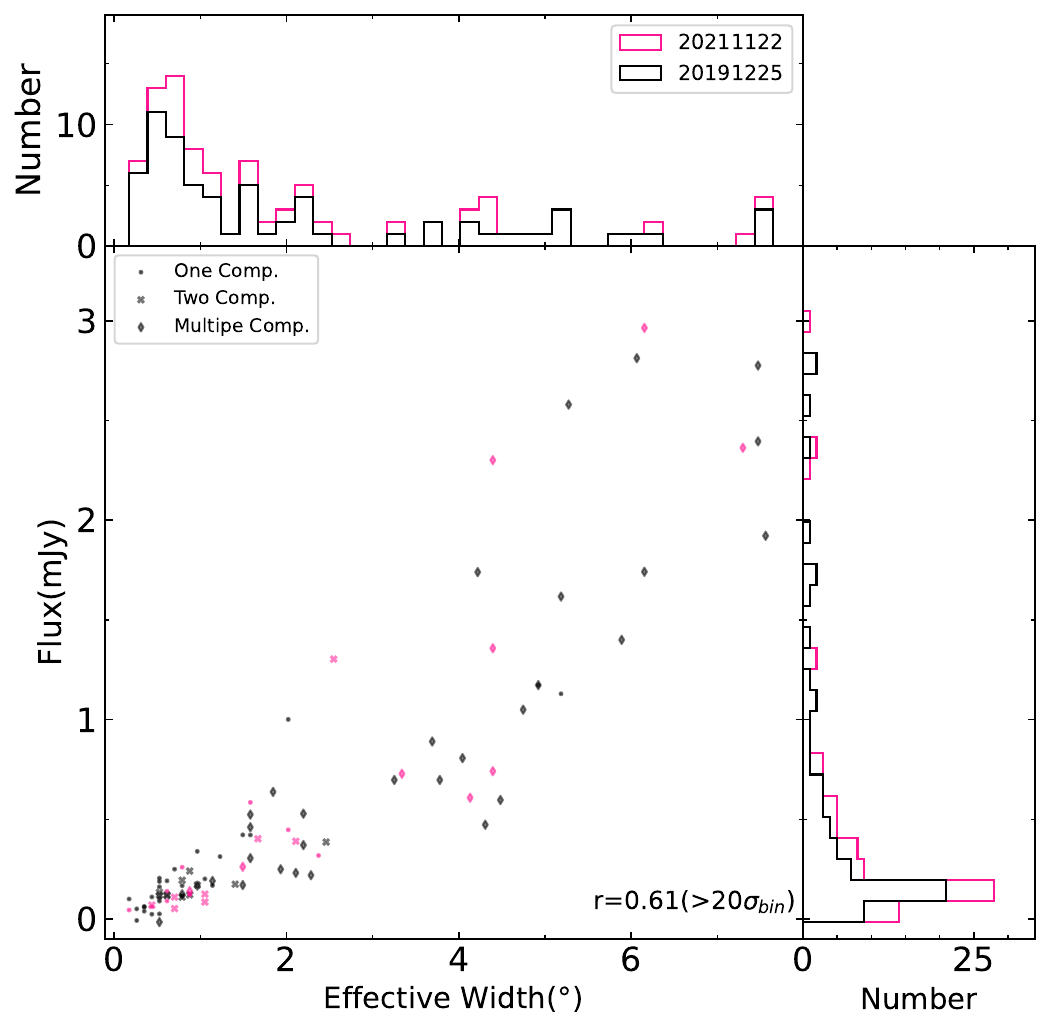}}
   \caption{Correlation between pulse intensity and profile width 319 pulses have peak intensities above 5$\sigma_{\rm bin}$, and 95 pulses above 20$\sigma_{\rm bin}$. The effective width represents the summed widths of all the phase bins above the 5$\sigma_{\rm bin}$ detection levels. The correlation coefficient is indicated in the bottom right corner of each panel. The single peak, double peaks, and multiple peaks are represented by dots, forks, and diamonds in left bottom panel of  plot. Observations on 20191225 and 20211122 are represented by black and pink dots.}
   \label{Fig:EffectiveW_F}
\end{figure*}
193 individual pulses exhibit a single profile peak from the two observations. Pulses are composed of one peak, that is typically narrow in width, as shown in Figure~\ref{Fig:Ex.Comp}(a). Their emission can be generated at the leading ($<$0$^\circ$), central ($\sim$ 0$^\circ$) or trailing ($>$0$^\circ$) part of the average pulse profiles. Histograms for their peak phases are demonstrated in Figure~\ref{Fig:OneComp} (a). They roughly follow a normal distribution with the average of $-0.18^\circ$, and the standard deviation of $1.2^\circ$. Their mean values are listed in column (3) of Table~\ref{tab:1}. More pulses tend to have small peak intensities, their number decays exponentially with the increase of peak intensity, as shown in Figure~\ref{Fig:OneComp} (b). The characteristic peak intensity is 19.6 mJy. The profile widths are measured from the left and right profile edges with intensity larger than $5\sigma_{\rm bin}$. Distribution of which also follows an exponential distribution with a characteristic width of $0.13^\circ$, as demonstrated in Figure~\ref{Fig:OneComp} (c). Their mean widths are listed in column (6) of Table~\ref{tab:1}. The fractional linear polarization of these individual pulses varies from about 0\% to 100\% with a mean of about 55\%. Only a few individual pulses have circular polarization with either positive or negative senses. Its maximum reaches 28\%. Two individual pulses exhibit orthogonal modes with jumps happening at phases of $-0.5^\circ$ and $2.0^\circ$. 

\subsubsection{Profiles with double peaks} 

61 individual pulses exhibit two profile peaks from the two observations. These pulses are compounded from two peaks, as shown in Figure~\ref{Fig:Ex.Comp}(b). Both peaks can be completely or partially resolved. They can also be located at the leading, central or trailing part of the average pulse profile. Their peak intensity varies from 16.68 to 413.4 mJy, and exponential distribution. Their widths are about 1.1$^\circ$ on average. The fractional linear polarization ranges from 0.7\% to 100\% with an average of about 55\%. The fractional circular polarization is generally low, the maximum of which is about 33\%. Their senses are diverse, some exhibit multiple sense reversals.
Three individual pulses exhibit orthogonal mode jumps at phases of $-2.0^\circ$ and $-0.2^\circ$. 

\subsubsection{Profiles with multiple peaks}

65 individual pulses exhibit multiple profile peaks from the two observations, as shown in Figure~\ref{Fig:Ex.Comp}(c). Their widths range from $0.48^\circ$ to $7.38^\circ$ with a mean of about $3.2^\circ$. They have an average peak density of about 150.7 mJy, and are generally stronger than those with single or double peaks. 10 profiles exhibit orthogonal mode jumps at phases of $-2.0
^\circ$, $-1.5^\circ$,$-0.2^\circ$, $0.2^\circ$, $2.1^\circ$ and $3.5^\circ$, respectively. The phase where it happens is consistent with that of the integrated pulse profile. Their fractional linear polarization is about 50\%. The circular polarization is dominated by right-handed polarization, some pulses even exhibit four sense reversals.

In summary, it is evident from Table~\ref{tab:1} and the analysis here that pulse profiles with more peaks tend to have larger width, peak intensity and flux density than those with fewer peaks. This property is further supported by the correlation between pulse width and its flux density in Figure~\ref{Fig:EffectiveW_F}. The correlation is partly due to the signal-to-noise ratio cut off. That is to say, brighter pulses tend to have more bins. When the threshold increases from 5$\sigma_{\rm bin}$ to 20$\sigma_{\rm bin}$, the influence of signal-to-noise ratio cut off is reduced.  Pulse width and its flux density remain correlated with a correlation coefficient of 0.61.
It is obvious from the figure that the burst emission of J1854+0306 is dominated by narrow (less than 1$^\circ$) and weak (less than 0.5 mJy) pulses, which are inter spaced by occasional wide and intense bursts. Moreover, orthogonal modes are present for all the profile types, but their jumps take place at different rotation phases.

\section{Discussion and conclusions}

In this paper, we investigate the polarized emission of RRAT J1854+0306 using highly sensitive observations from FAST. Its emission is dominated by nulls with a nulling fraction of about 53.2\%, which is inter spaced by narrow and weak pulses with occasional wide and intense bursts. The individual pulses exhibit diverse profile morphology, exhibiting single, double and multiple peaks. The intensity and width of the pulses are correlated, and bright pulses generally have wide profiles with multi-peaks. Its burst emission is typically of one rotation. Occasionally, the bursts last for two or three or even 5 rotations at the most. Moreover, the burst emissions exhibit no significant periodicity.

These pulse exhibit diverse polarization behaviours. Their degree of linear polarization can reach 100\% for some pulses, and their circular polarization exhibits various senses and variation. These features are related to the density distribution of relativistic particles and their emission processes \citep[e.g.][]{Wang+etal+2012} and/or caused by the propagation effects \citep[e.g.][]{BP12, Wang+etal+2014}. For some pulses, the position angles depart a lot from the average one, which might be caused by emission generated from different plasma conditions of the magnetosphere \citep[e.g.][]{BCW91,gan+05,Wang+etal+2012}. The geometry parameters are estimated from the `S' shaped PA, which demonstrates that its emission beam is cut by sight lines with small impact angles. The polarization of its individual pulses exhibits two distinct orthogonal modes, O-mode and X-mode. According to the pulsar emission theory accounting for emission processes and propagation effects within pulsar magnetosphere \citep{Wang+etal+2014}, the X-mode emission dominates the central part of pulse profile and the O-mode emission to the leading and trailing parts, when the emission beam is cut by small impact angels. The observational facts that superposition of both modes leads to depolarization and twice jumps of position angles also meet with the predictions by \citet{Wang+etal+2014} for the typical pulsar magnetosphere conditions. Hence, the behaviour of polarized emission of RRAT J1854+0306 demonstrates that its emission originates from a magnetosphere similar to those of normal pulsars.

\section*{Acknowledgements}

This work made use of the data from FAST (Five-hundred-meter Aperture Spherical radio Telescope).  FAST is a Chinese national mega-science facility, operated by National Astronomical Observatories, Chinese Academy of Sciences. P.F.W. is supported by the National Key R\&D Program of China (No. 2021YFA1600401 and  2021YFA1600400), National Natural Science Foundation of China (NSFC, grant No. 12133004) and the National SKA program of China (No. 2020SKA0120200). M.Z.K. is supported by the Astronomical Union Foundation under grant No. U1831126, the Natural Science Foundation of Hebei Province No. A2019205100 and the National SKA Program of China (Grant No. 2022SKA0120103).

\section*{Data Availability}

Original FAST observational data are open sources in the FAST
Data Center according to the FAST data 1-year protection policy. The folded and calibrated data in this manuscript can be obtained from authors by reasonable requests.



\bibliographystyle{mnras}
\bibliography{text} 


\bsp	
\label{lastpage}
\end{document}